# Randomness Evaluation of a Genetic Algorithm for Image Encryption: A Signal Processing Approach

Zoubir Hamici, *Senior Member, IEEE*

**Abstract**— In this paper a randomness evaluation of a block cipher for secure image communication is presented. The GFHT cipher is a genetic algorithm, that combines gene fusion (GF) and horizontal gene transfer (HGT) both inspired from antibiotic resistance in bacteria. The symmetric encryption key is generated by four pairs of chromosomes with multi-layer random sequences. The encryption starts by a GF of the principal key-agent in a single block, then HGT performs obfuscation where the genes are pixels and the chromosomes are the rows and columns. A Salt extracted from the image hash-value is used to implement one-time pad (OTP) scheme, hence a modification of one pixel generates a different encryption key without changing the main passphrase or key. Therefore, an extreme avalanche effect of 99% is achieved. Randomness evaluation based on random matrix theory, power spectral density, avalanche effect, 2D autocorrelation, pixels randomness tests and chi-square hypotheses testing show that encrypted images adopt the statistical behavior of uniform white noise; hence validating the theoretical model by experimental results. Moreover, performance comparison with chaos-genetic ciphers shows the merit of the GFHT algorithm.

## I. Introduction

SECURITY has become a key issue in the virtual world of modern life due to the global connectivity which has crucial implications on sensitive information being within reach of possible remote malicious entities. Therefore, the viability of the information society relies upon a high degree of security insured by the deployment of novel algorithms in computational and artificial intelligence reinforced by molecular biology inspired schemes. In fact, advances in artificial intelligence towards automated cryptanalysis together with the continuous evolution and increased computing power endanger secure communications that rely heavily upon algorithms and their resistance to cryptanalysis attacks. On the other hand, the state of the art cryptographic algorithms surpassed the performance of a mathematical structured algorithms [1], by exploiting natural complexity upon which is based genetic structures of biological organisms and their evolution [2], [3]. Indeed, novel algorithms are required to overcome the rapid growth of computational power and intelligence which makes modern cryptographic algorithms keep getting broken.

Prof. Zoubir Hamici is with the Department of Electrical Engineering, Faculty of Engineering and Technology, Al-Zaytoonah University of Jordan (ZUJ), Amman 11733, Jordan (e-mail: zhamici@gmail.com). This work was partially funded by Al-Zaytoonah University of Jordan, deanship for scientific research. Grant No. 18/18/2018-2019.

The continuous spread of digital images; from military, medical, commercial to private usage, makes them a cornerstone in modern digital communication. Moreover, images are also excellent media for covert communication by steganography [4]-[7]. Image encryption finds its application alongside with steganography for data hiding in encrypted images. In [8], authors propose to consider the patch-level sparse representation, with the objective of better exploring the correlation between neighbor pixels where the leading residual errors are encoded and self-embedded within the cover image. In [9] a private key cryptosystem based on the finite-field wavelet is presented. The encryption and decryption are performed by the synthesis and analysis banks of the nonlinear finite-field wavelet transform; whose filter coefficients are determined by the keys of the users. Under the assumption of plaintext attacks, the security of a diffusion mechanism used as the core cryptographic primitive in some image cryptosystems based on chaotic orbits and quantum walks is investigated in [10]. The proposed analysis is validated by numerical examples. Moreover, the importance and feasibility of applying a joint signal processing and cryptographic approach to multimedia encryption is discussed in [11].

The rapid growing in bio-inspired computation opens the doors to use its natural complexity in cryptographic field. The DNA computing cryptography along with chaos theory are an emerging and very promising direction in cryptography research. In [12] authors proposed an image cipher with combined permutation and diffusion; first, the image is partitioned into blocks, then a spatiotemporal chaos is employed to shuffle the blocks and change the pixels' values. Wang et al. presented a scheme based on DNA sequence operations and a chaotic system; in first phase, they perform bitwise XOR operation on the pixels of the plain image by the spatiotemporal chaos system, then, in a second phase a matrix is obtained by encoding the confused image using a DNA encoding rule and finally, a new initial condition of a coupled map lattice is generated according to the DNA matrix [13]. In [14] authors presented a DNA encryption based on one-time pads. They detailed the procedure for two one-time-pads encryption schemes where a substitution method uses libraries of distinct pads, each of which defines a specific, randomly generated, pair-wise mapping; and an XOR scheme utilizing molecular computation with indexed random key strings. Genetic cryptography has also been used for ad-hoc networks, secure data transfer or protecting multimedia information [15]. Many research works have been performed on combining chaos theory with DNA sequencing for the purpose of image encryption with results comparison of several two-dimensional chaotic maps [16]-[19].

Regarding security analysis based on randomness evaluation, authors in [20] studied the discrete Fourier transform (DFT) randomness test NIST-SP-800-22 released by the National Institute of Standards and Technology. They proved that a power spectrum, which is a component of the test statistic, follows a chi-squared distribution, the effectiveness of their approach was verified by experimental results. Other works dealing with image encryption investigate reaction-diffusion neural networks application in secure image communications [21] and combining secret image sharing with image encryption [22]. In [23] security evaluations of a class of randomized encryption schemes are performed employing an information-theoretic approach.

This paper presents randomness evaluation of the GFHT algorithm for image communication with OTP scheme. A preliminary introduction of the present GFHT algorithm was reported in [24] while an application in biomedical wireless sensor network gateway with simulation and experimental results was presented in [25], where the GFHT algorithm out-performs the most widely used conventional ciphers in avalanche effect and was ranked second in terms of speed, after AES algorithm. The signal processing approach for security analysis presented hereinafter, is based on random matrix theory circular law, power spectral density, avalanche effect, 2-D autocorrelation, pixels randomness tests and chi-square, $\chi^2$, hypotheses testing. Moreover, performance comparison with some genetic and chaos theory algorithms is presented. Beside its genetic scheme, the GFHT offers the feature of an OTP since every image is encrypted with different key, while keeping the main passphrase unchanged. The paper is organized as follows: Section II, presents the GFHT algorithm and its model. Section III carries out the cipher randomness analysis showing mathematical foundations on randomness of encrypted images. In section IV comprehensive results analysis of the GFHT algorithm is provided reinforced by extended comparison with chaos and genetic algorithms. Finally, a conclusion on the present contribution is provided in section V.

## II. GENETIC ALGORITHM: KEYS GENERATION AND ENCRYPTION PROCESS

The GFHT algorithm [24] is divided into two stages; the OTP keys generation process and a round-based encryption process. For the Keys Generation, the same passphrase generates different key for each encrypted image. The principal key and sub-keys $V\_Key$, $H\_Key$ are given.

$$P\_Key = V\_Key^T * H\_Key \ (mod \ 256) \quad (1)$$

$$V\_Key = [a_1 \ \ldots \ a_i \ \ldots \ a_N] \quad (2)$$

$$H\_Key = [b_1 \ \ldots \ b_j \ \ldots \ b_M] \quad (3)$$

For Gene Fusion and Horizontal Gene Transfer, the principal key, $P\_Key$, will act as a pathogen, by introducing gene fusion into the image. The $R$ layer key is the same principal key; the $G$ layer key is obtained by a 3 bits' reversal (bases reversal) and the $B$ layer key is a 6 bits' reversal as given by (4).

$$P\_Key = \begin{cases} P_{Key0} \rightarrow 0bits \otimes & for \ R \\ P_{Key3} \rightarrow 3bits \otimes & for \ G \\ P_{Key6} \rightarrow 6bits \otimes & for \ B \end{cases} \quad (4)$$

A genetic crossover operator ($\otimes$) defines the permutation of rows of a matrix $I$ as: $n^{(l)} \leftarrow V\_Key(m^{(l-1)})$ and the permutation of columns as: $q^{(l)} \leftarrow H\_Key(p^{(l-1)})$. The full crossover operator of an image $I$ at round $l$, is defined by (5). The symbol (:) indicates an unchanged state of the corresponding row or column.

$$\otimes \{I\}^{(l)}(n,q) \stackrel{def}{=} \begin{cases} I^{(l)}(m,:) \leftarrow I^{(l-1)}(V\_Key(m),:) & 1 \leq m \leq M \\ I^{(l)}(:,p) \leftarrow I^{(l-1)}(:,H\_Key(p)) & 1 \leq p \leq N \end{cases} \quad (5)$$

The multi-round recursive encryption process at round $l$, is defined by (6).

$$I^{(l)} = \begin{pmatrix} R^{(l)} \\ G^{(l)} \\ B^{(l)} \end{pmatrix} = \otimes \begin{cases} R^{(l-1)} \oplus P_{Key0} \\ G^{(l-1)} \oplus P_{Key3} \\ B^{(l-1)} \oplus P_{Key6} \end{cases}^{(l-1)} \quad 1 \leq l \leq L \quad (6)$$

The nonlinear mapping permutation keys are used to swap rows and columns at round $l$, using (7) and (8) given below:

$$n^{(l)} = V\_Key(m^{(l-1)}) \quad (7)$$

$$q^{(l)} = H\_Key(p^{(l-1)}) \quad (8)$$

The pixels of $I_R$, $I_G$ and $I_B$ are encrypted recursively from the round 0 of unencrypted pixels to round $L$ of fully encrypted pixels. Each pixel is encrypted, moved, re-encrypted with different values in the $P\_Key$ matrix, where positions are controlled with the $V\_Key$ and $H\_Key$ sub-keys. Equations (9), (10) and (11) describe the process. The superscripts indicate the round number. Each RGB layer is encrypted by its $P\_Key$ matrix. Values of $n$ and $q$ are given by (7) and (8).

$$I_R^{(L)}(m,p) = \otimes \{ \ldots \otimes \{\otimes \{I_R^{(0)}(m,p) \oplus P_{Key_0}\}(n^{(1)},q^{(1)}) \oplus \ldots \\ P_{Key_0}\}(n^{(2)},q^{(2)}) \oplus \ldots \oplus P_{Key_0}\}(n^{(L)},q^{(L)}) \quad (9)$$

$$I_G^{(L)}(m,p) = \otimes \{ \ldots \otimes \{\otimes \{I_G^{(0)}(m,p) \oplus P_{Key_3}\}(n^{(1)},q^{(1)}) \oplus \ldots \\ P_{Key_3}\}(n^{(2)},q^{(2)}) \oplus \ldots \oplus P_{Key_3}\}(n^{(L)},q^{(L)}) \quad (10)$$

$$I_B^{(L)}(m,p) = \otimes \{ \ldots \otimes \{\otimes \{I_B^{(0)}(m,p) \oplus P_{Key_6}\}(n^{(1)},q^{(1)}) \oplus \ldots \\ P_{Key_6}\}(n^{(2)},q^{(2)}) \oplus \ldots \oplus P_{Key_6}\}(n^{(L)},q^{(L)}) \quad (11)$$

## III. ALGORITHM SECURITY ANALYSIS

Randomness analysis of the GFHT algorithm is based on well-established metrics in image cryptography and random matrix theory presented in the following sub-sections

### 3.1 Pixels Randomness Tests with NPCR and UACI

In order to provide a metric for differential pixels change, the Number of Pixels Change Rate (NPCR) and the Unified Average Changing Intensity (UACI) are adopted for describing the ability to resist the differential attack [29]. The *NPCR* metric is defined by:

$$NPCR(\%) = \frac{\sum_{i,j} D(i,j)}{M.N} \times 100 \quad (12)$$

with pixels' difference:

$$D(i,j) = \begin{cases} 0 & if \ I_C(i,j) = I_{C1px}(i,j) \\ 1 & Otherwise \end{cases} \quad (13)$$

where *M* and *N* are the height and width of the image, respectively. $I_c(i,j)$ and $I_{c1px}(i,j)$ are the pixels in a cipher-image and the cipher-image after changing one pixel in the plaintext image respectively. In that regard a number of images are used for testing, in the results section, in order to obtain an average value of *NPCR*, where an extensive differential cryptanalysis is presented. Also, the *UACI* has been used to quantify the differential analysis between two images enciphered with random pixels change with an ideal value expected to oscillate around the value of 1/3 (see Appendix). The *UACI* is given by

$$UACI(\%) = \frac{1}{N \times M} \sum_{i,j} \frac{|C_1(i,j) - C_2(i,j)|}{255} \times 100 \quad (14)$$

where *M*, *N*, are the length and width of each image. $C_1$ and $C_2$ are the enciphered images with random pixels change.

*3.2 Power Spectral Density*

A powerful signal processing space is the frequency domain where cipher randomness behavior is carried out by the analysis of the power spectral density of the process being considered. If the encryption process is random, then a constant density (dB/Hz) should be observed on the signal spectrum. Let *x(n)* be a discrete wide-sense stationary random process representing the encrypted data with autocorrelation function given by

$$r_{xx}[n, n-l] = r_{x[n]}[l] = \mathbb{E}[x[n]x^*[n-l]] \quad (15)$$

where $\mathbb{E}(.)$ is the expectation function and the star superscript $(.)^*$ denotes the complex conjugate. Remember that a white noise $x[n]$ has its expected value $\mathbb{E}[x[n]] = 0$, and recall that independent variables have the autocorrelation $r_{xx}[l] = \mathbb{E}[x[n]x[n-l]]$ for real signals and since the random samples inside $x[n]$ are independent, then the autocorrelation is the product of the individual expected values $r_{xx}[l] = \mathbb{E}[x[n]].\mathbb{E}[x[n-l]]$ which concludes to:

$$r_{xx}[l] = \begin{cases} \mathbb{E}[x[n]]\mathbb{E}[x[n-l]] & \text{if } l \neq 0 \\ \mathbb{E}[x^2[n]] & \text{if } l = 0 \end{cases} \quad (16)$$

thus, the following result, when verified, proves the pure randomness of a process: $r_{xx}[l] = \sigma^2 \delta[l]$, where $\sigma^2$ is the variance of the analyzed process and $\delta$ is Dirac delta function with $\delta[l] = 1$ for $l = 0$ and $\delta[l] = 0$ otherwise. Hence a null autocorrelation is the result of a pure random output, validating the security of the cipher scheme as it will be shown in the results section for both 1-D and 2-D. In the frequency domain, for each sample function of the process $x[n]$, the truncated Discrete Fourier Transform, over a time slot *T*, is given by

$$X_T[k] = \sum_{n=0}^{N-1} x[n]e^{-j2\pi kn/N} \quad (17)$$

where *k, n, N*, are the discrete frequency, discrete time and the number of samples respectively. The corresponding truncated power spectral density is given by

$$PSD_T[k] = \frac{1}{N}|X_N[k]|^2 \quad (18)$$

since $x[n]$ is a random process, for each discrete frequency, *k*, $PSD_T[k]$ is also a random univariate. Let us denote the expectation of truncated power spectral density by

$$PSD_{x[n],N}[k] \stackrel{\text{def}}{=} \mathbb{E}\left[\frac{1}{N}|X_N[k]|^2\right] \quad (19)$$

and therefore, the power spectral density of the process is defined by

$$PSD_{x[n]}[k] = \lim_{N \to \infty} S_{x[n],N}[k] \quad (20)$$

The Wiener-Khintchin theorem states that the limit given by (20) exists for all discrete frequencies an its value for the discrete random process is given by the DFT of the autocorrelation function as

$$PSD_{x[n]}[k] = \sum_{l=-N}^{N} r_{x[n]}[l] e^{-j2\pi kl/N} \quad (21)$$

For a random process, the autocorrelation function shows an impulse at the origin (time delay *l=0*), which will be illustrated by a constant (flat) power spectral density over all the frequency range. Illustrations will be shown in section IV, subsection 4.2

*3.3 2-D Auto-Correlation Analysis*

The normalized auto-correlation coefficient, *ρ*, also known as Pearson's Correlation Coefficient (PCC), measures the strength of a linear relationship between two vectors or arrays; the value of *ρ* always spans the interval [-1,1]. Consequently, in order to measure the dependability between pixels inside an image denoted by a matrix *X*, we perform three directional auto-correlation measures on *X*; horizontal, vertical and diagonal auto-correlations. Let $X_{i,j}$ be the reference matrix, to be correlated to a horizontally one-pixel shifted matrix, $Y=X_{i+1,j}$. The vertical one-pixel shifted matrix is defined by $Y=X_{i,j+1}$. The diagonal shifted matrix is defined by $Y=X_{i+1, j+1}$ (Fig.2). Three correlations are performed in order to measure the dependability of pixels. The covariance of two matrices *X* and *Y* normalized over their inter-energy, since the norm of a signal is defined as the square root of its total energy which provides the normalized correlation coefficient, *ρ*.

$$\rho_{X,Y} = \frac{\text{cov}(X,Y)}{\sigma_X \sigma_Y} \quad (22)$$

where *cov(.)* is the covariance function, $\sigma_X$ is the standard deviation of the image *X* and $\sigma_Y$ is the standard deviation of the image *Y*. The covariance is given based on the Expectation and the mean values of the two images.

$$\text{cov}(X,Y) = \mathbb{E}[(X - \mu_X)(Y - \mu_Y)] \quad (23)$$

where $\mu_X$ is the mean of the image X, and $\mu_Y$ is the mean of the image Y. Data is shifted by subtracting their respective means in order to have an average of zero.

$$\rho_{X,Y} = \frac{\sum_{i,j}(X_{i,j} - \mu_X)(Y_{i,j} - \mu_Y)}{\sqrt{\sum_{i,j}(X_{i,j} - \mu_X)^2}\sqrt{\sum_{i,j}(Y_{i,j} - \mu_Y)^2}} \quad (24)$$

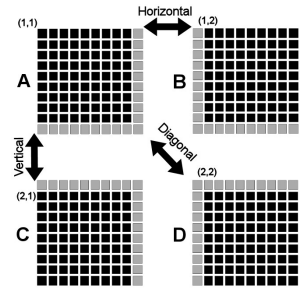

Fig. 1. Directional autocorrelation. A-B: Horizontal autocorrelation, A-C: vertical autocorrelation and A-D: diagonal autocorrelation.

The analysis of the correlation coefficient will reveal the capability of the cipher to remove dependencies between adjacent pixels. A meaningful image always has a high correlation coefficient; practically in the range [0.7-1.0] depending on the class of image; natural scene, medical, synthetic etc. However, an ideal cipher should verify the null hypothesis; practically producing a coefficient that is less than 0.01 as illustrated in the results section. Figure 1 shows how directional auto-correlation is performed; Matrix A is the original image (layer) reduced by one pixel in each direction, black squares are the pixels to be correlated while grey square are ignored pixels. Horizontal correlation is between A and B, vertical correlation is between A and C, and diagonal correlation is between A and D.

## IV. RESULTS AND DISCUSSION

In this section, results to evaluate the performance of the proposed algorithm are provided. Results are obtained with personal computer equipped with intel i7-4510U dual core processor and 16 GB of RAM running at 2.0 GHz under windows 10 pro 64-bit operating system. To demonstrate the performance and effectiveness of the proposed cipher, results of different metrics are shown in sub-sections: 4.1 to 4.5.

### 4.1 Random Matrix Theory

An ideal RGB encrypted image is manifested by the behavior of the three layers acquiring the properties of random matrices, therefore, random matrix theory (RMT) is used to verify the effect of the genetic algorithm in producing random matrices. After, introducing RMT model for eigenvalues spectral distribution, the best way to validate this expectation is by experiments on RGB images. This is an effective way to contemplate the occurrence of convergence of the spectral distribution in encrypted images as random matrices outputs.

Let $X = X_N(i,j)$, $1 \leq i,j \leq N$ be a random matrix of size N×N, with eigenvalues vector $\lambda = [\lambda_1 \cdots \lambda_N]$. A normalized random matrix $X$, $X = \{R, G, \text{or } B\}$, is defined by (25)

$$X_n = \frac{1}{\sqrt{N}} X_N(i,j), 1 \leq i,j \leq N \qquad (25)$$

with zero mean $\mathbb{E}[X_n] = 0$ and a unit variance $\mathbb{E}[X_n^2] = 1$. The eigenvalues spectral distribution is defined by (26)

$$G_n(x,y) = \frac{1}{N}\sum_{k=1}^{N} I_{\{\text{Re}\{\lambda_k\} \leq x, \text{ Im}\{\lambda_k\} \leq y\}} \qquad (26)$$

where $I_{\{.\}}$ denotes the occurrence of an event and the expectation of the spectral distribution function, $\mathbb{E}[G_n(x,y)]$, converges to a uniform distribution in the unit disk in $\mathbb{R}^2$ [30].

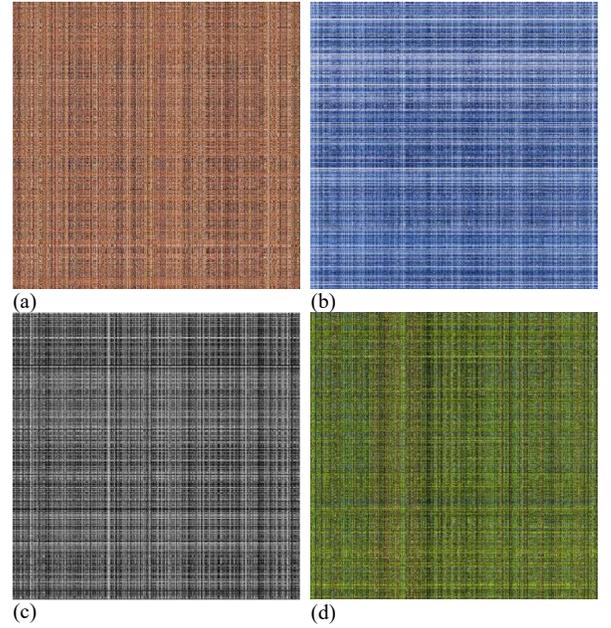

Fig. 3. HGT for the four test images of Fig. 2.

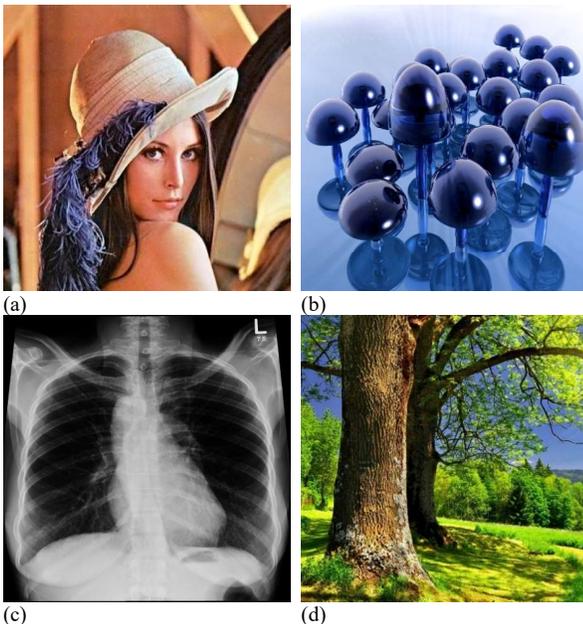

(a) (b) (c) (d)
Fig. 2. Original test images. (a) Lena-1024, (b) Mushrooms, (c) X-ray, (d) Nature.

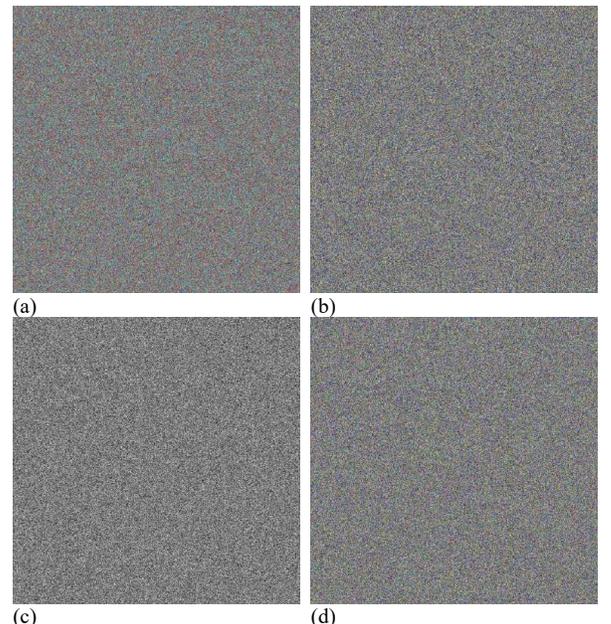

Fig. 4. Encryption using GF and HGT for images of Fig. 2.

Figure 3 shows HGT results on images in Fig.2, while Fig. 4 shows the fully encrypted images. All RGB layers of the test images show uniform eigenvalues distribution inside a unit circle. This is illustrated in Fig. 5 by showing the eigenvalues distribution of both unencrypted and encrypted RGB layers of "Lena 400". This uniform distribution carries out that the algorithm produces random matrices conform with RMT. The experimental testing of unencrypted and encrypted images concluded about the behavior of random outputs of the algorithm. In fact, while the eigenvalues of images in Fig.2 are randomly distributed in a two-dimensional complex space defined by the imaginary parts versus real parts, mainly concentrated around the origin (Fig.5- d, e, f), the normalized encrypted images in Fig.4 have a uniform distribution inside a unit circle (Fig.5- a, b, c). These results have decisive inference on encryption algorithm producing image layers following RMT circular law rules. The dyadic product *P_Key* in its three layers given by (5) produces three RGB encrypted layers, hence each layer is encrypted with a different version of the principal key.

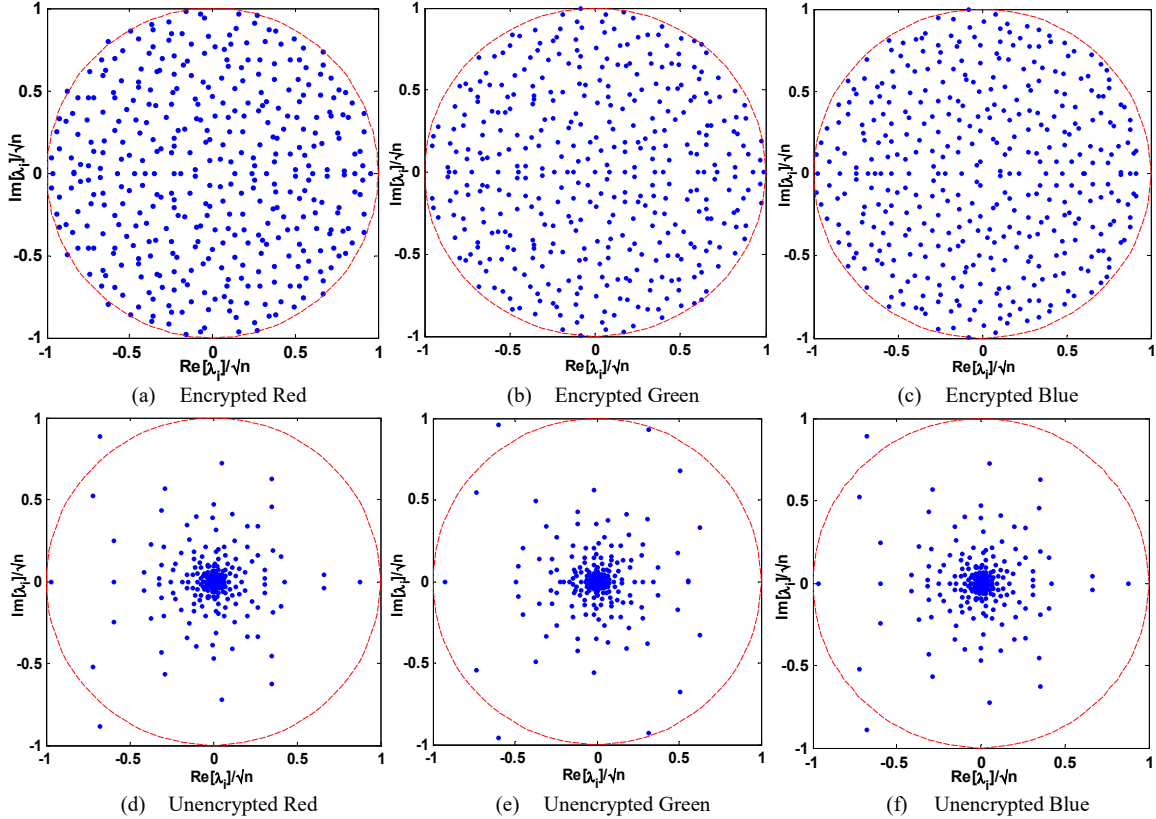

Fig. 5. Eigenvalues spectral distribution. Encrypted RGB layers spectral distribution (a, b, c) and their corresponding unencrypted counterparts (d, e, f).

In terms of computation speed GF is performed in 8ms per round (three layers) for an image of 1024×1024, and HGT is achieved in about 60ms. Figure 6 illustrates the eigenvalues distribution inside the unit circle. The R, G and B eigenvalues of the encrypted matrices coincide with the theoretical uniform normalized area distribution.

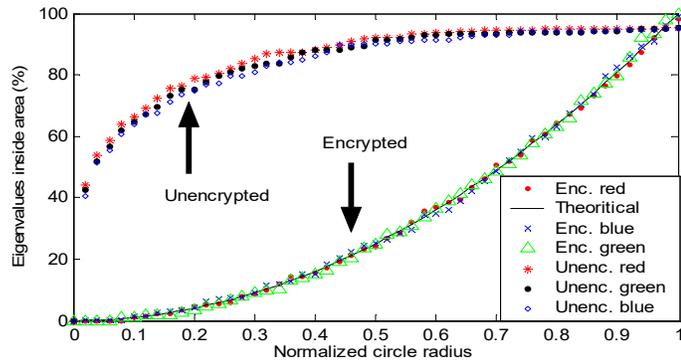

Fig. 6. Uniform spectral distribution of eigenvalues coinciding with the theoretical uniform area distribution of a unit circle (Area=π.radius$^2$).

The unencrypted layers have more than 75% of eigenvalues distributed inside 20% of the unit circle radius.

### 4.2 Power Spectral Density

A second characteristic illustrating the encrypted image as random process, is shown by a constant Power Spectral Density (PSD). The PSD is a metric representing the distribution of the signal average power as a function of spatial frequency, it is the average of the squared magnitude of the truncated Discrete Fourier Transform. The PSD is also obtained by the Fourier Transform of the autocorrelation function of the discrete Wide-Sense Stationary (WSS) random process. Figure 7 depicts the flat envelope of the two-dimensional PSD for an image size 1024×1024 pixels obtained by the two-dimensional Discrete Fourier Transform, the impulse at the 2D DFT represents the DC component of the spatial frequency space. Figure 8 illustrates the PSD estimate using the Welch spectral density estimation showing the spectrum from DC up to the maximum spatial frequency.

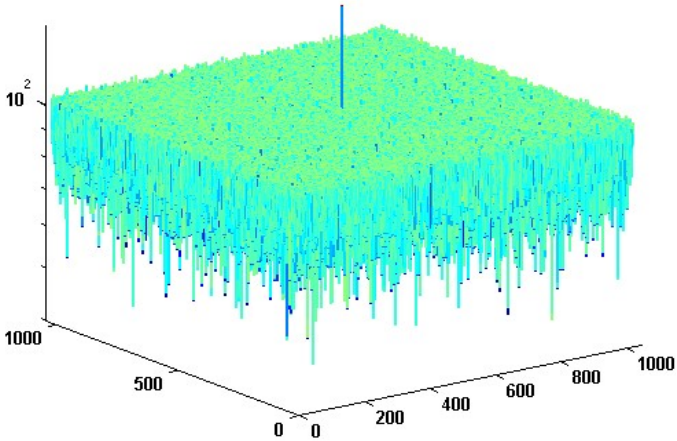

Fig. 7. Power spectral density as obtained by 2D DFT- RGB merged layers).

The Welch periodogram uses a sliding window of 1024 pixels and an overlap of 50%. The impulse at DC is the mean of the encrypted data. The encryption algorithm is generating data that is characterized as a white noise illustrated by a flat power spectral density across the full spatial frequency range. Figure 8 illustrates the behavior of encrypted image RGB layers compared with white noise. The Welch PSD estimate of the encrypted image layers coincides with the PSD of a white noise at -25.7 dB/[Hz]. The wavering (ripples) of the encrypted image are within 0.15 dB for all RGB layers and white noise as shown by Fig.8 legend.

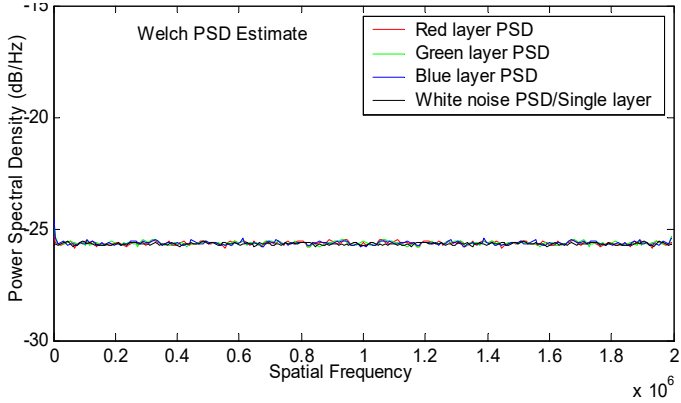

Fig. 8. Welch periodogram of the power spectral density (PSD in dB/[Hz]) of encrypted data. PSD for image size of 1Megabyte.

### 4.3 Avalanche Effect and Pixels Randomness Tests

Regarding the avalanche effect property, Table I shows the analysis of four images for the avalanche effect of the algorithm produced by the nullification of one arbitrary pixel in each test image. For each case, two keys (Keys: *Key A* and *Key B*), are generated by the algorithm using the same passphrase (password). Each key is obtained by the hash of the passphrase concatenated with le last 32bits of the hash of the image (salt). The metrics for pixels randomness tests are given by the *NPCR* and *UACI*. Figure 9 shows NPCR for 100 tests performed on RGB layers of twenty images, each one encrypted with different keys. The change of the salt produces an avalanche effect for the encrypted images with an average number of pixels change rate (*NPCR %*) of 99%. Analysis of the *UACI* is given in Fig.10 for 100 tests on the RGB layers which provides an illustration of the validity of the cipher scheme with an average *UACI* of 33.45%. The latter value represents an accuracy of 99.64% relative to the ideal value. Despite the fact that the algorithm structure requires the exchange of a partial-hash value of the data block as a Salt along with the ciphertext, a brute force attack on the partial-hash value will not be feasible since it is only a portion in the reconstruction of the main key. Besides that, the algorithm structure protects the general key, since no direct hash is used or transmitted. Moreover, for the data partial hash, there is no known way to recover a full hash-value from one part of it.

### 4.4 Scattering and Pearson Correlation Coefficient

In order to observe the 2-D correlation, scatter images are useful for interpreting the trends in statistical data. Each observation (or point) in a scatter image has two coordinates; the first corresponds to the intensity of a pixel in the reference image while the second coordinate corresponds to the intensity of the corresponding adjacent pixel in the shifted image (horizontal, vertical or diagonal). The point representing that observation is placed at the intersection of the two coordinates. If the scattered data show a positive slope pattern as we move from left to right, this indicates a positive relationship between *X* and *Y*; an increase in one variable produces the increase of the corresponding variable. If the scattered data do not seem to resemble any kind of pattern, then no relationship exists between *X* and *Y* and the analysis concludes about the pure randomness of the analyzed data. One pattern of special interest is the linear pattern, where the scattered data produces a cloud of points around the diagonal line; a cloud of points that has a general look of a line going uphill. Figure 11 (a, b, c), show the scatter points around the positive diagonal for the red, green and blue layers respectively (2D diagonal correlation).

TABLE I: NUMBER OF PIXELS CHANGE RATE: THE PROPERTY OF AVALANCHE EFFECT PRODUCED BY A NULLIFICATION OF ONE ARBITRARY PIXEL IN EACH IMAGE OF FIG.2 (4 IMAGES)

| # | Keys | Key (Hash 256bit- SHA256) | NPCR% |
|---|------|---------------------------|-------|
| a | Key A | 1F2C065079577EBC4E688DCE5A6D19F56F2A3419B3B1D272B0B4E9D2631D78B2 | 98.8 |
|   | Key B | 488ABEE73637A70FA2593199764B1DA901E22ABB32EEF63DD0F9DFAE09CA4E23 |  |
| b | Key A | 8F327415F2621DF3C76488EA7D170942FE5AB3BB94770078D935B08F6B8E913B | 96.4 |
|   | Key B | 2FDE6840D2FAC6CAF6591B22521AA666E2183A58677FF694FBDE69B1318560F4 |  |
| c | Key A | 42FE95556A24B1400B7BBC9D908D2FC25067BE25AE4E611909AFEEF63A6C525B | 97.4 |
|   | Key B | 568BDB7629E428E6EF5655AEF27FCDFC5349814174CD31E46A3193CE8754A03D |  |
| d | Key A | 38595343C782167EF725AC07BF070BA2D905079753F382172FD3EFF60308276A | 99.1 |
|   | Key B | 9D88B446CBAB9DF39AF7270B412C9122CB3E4BD2A12D844E7E1A8CE2FD3639B4 |  |

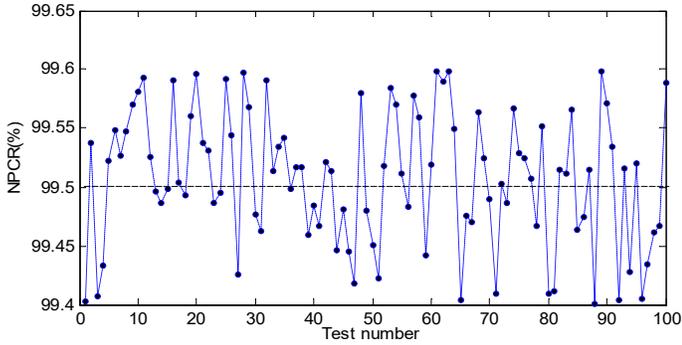

Fig. 9. Variation of the NPCR for 100 tests. The average NPCR for 100 tests is 99.50 %.

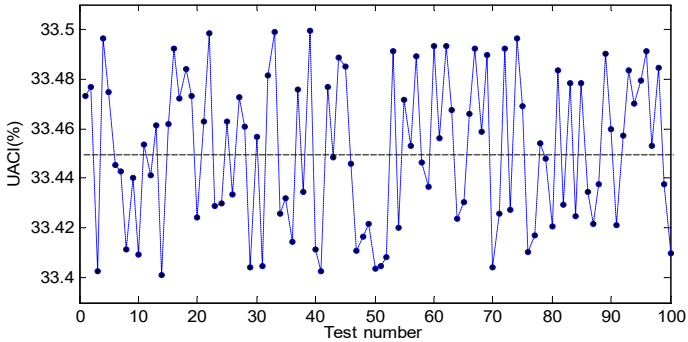

Fig. 10. UACI randomness test for red, green and blue layers.

The RGB layers are analyzed for one-pixel shift autocorrelation in horizontal, vertical and diagonal directions. These layers are from the Fig.2 (b) Mushroom image chosen since it represents the wide class of synthetic images. The images are reduced to 256×256 to show granularity in the scattering images. The correlation between adjacent pixels provides a high PCC for meaningful images since the values of adjacent pixels are among a meaningful shape, pattern or color.

An ideal cipher behavior should produce a cipher image having no correlation with adjacent pixels; hence it will serve as null Hypotheses testing. The correlation between a plain image and the adjacent or shifted counterpart will always produce scattered images that have high PCC, while the scattered analysis of cipher-images results in a pure noise image with no points cloud with a null PCC as shown in figures 11 (d, e, f). The PCC values for horizontally, vertically and diagonal directions are computed and shown in tables II and III for the RGB layers of plain images and cipher images. Further quantification and qualification analysis of images regarding pixel's dependability may exploit the directional scattering intensity, scattered-data variance, and expectation values, in addition to the PCC. Statistical inference based on PCC before and after encryption is aimed to test the null hypothesis of the PCC for all RGB layers and with all directional shifts (horizontal, vertical and diagonal). Furthermore, The PCC obtained from a directional autocorrelation may serve as a fast computational intelligence or a methodology of understanding to address the presence of meaningful information inside a complex image of real-world problems to which mathematical or conventional modeling can be useless due the complexity of required mathematical reasoning algorithms. Therefore, the PCC which transforms data extracted from images into certain commonly understood parameters, acts as a descriptor, that makes subsequent decisions and classifications. The PCC descriptor is used in Table II to analyze the directional dependability of pixels' intensity for the test images. Actually, the four test images represent four categories or classes of images; where "Lena-1024" represents the class of human photos, "Mushroom" represents the class of synthetic or computer made graphics, "X-Ray" represents the category of medical images such as X-ray [31], Medical Resonance Imaging (MRI) or Computerized Tomography (CT) and finally, "Nature" represents the class of natural views images. Among these four classes, "Nature" images have the lowest PCC while synthetic graphics have the highest PCC values. The maximum PCC values are shown in bold-font face in Table II while high PCC values appear in images that contain artificial objects or contain some regularities of color intensities in adjacent pixels. The PCC of the autocorrelation is greater than 0.77 for all images. The null hypothesis is tested and verified with the PCC of encrypted images shown in Table III where all values are less than 0.0075 carrying out the fact that the cipher removes dependability between adjacent pixels.

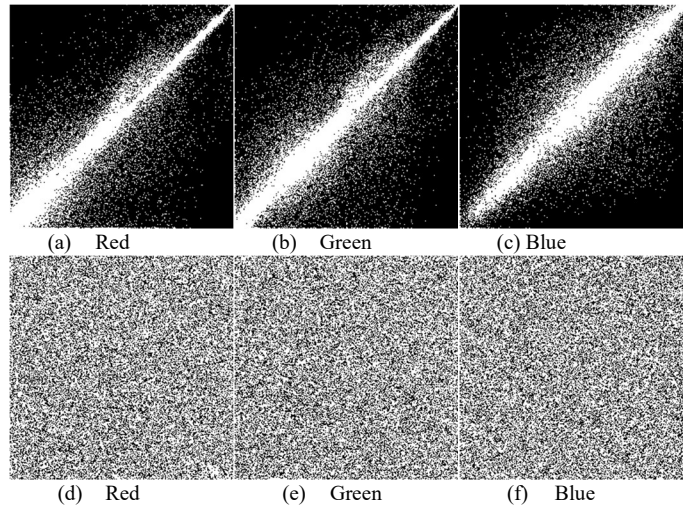

Fig. 11. Scattering images for diagonal autocorrelation for the red, green and blue layers (a, b, c) and their corresponding encrypted counterparts (d, e, f)

A comparison of the proposed cipher performance with some DNA and chaos-theory based ciphers presented hereinafter uses a well-known test greyscale image. An RGB color image is built of three independent stacked color channels (layers), each of which representing value levels of the given channel; red, green and blue primary color components. For a comparison purpose with previous works results using DNA and chaos theory logistic maps cryptography [12], [13], [16], [19], the blue layer of "Lena-1024" is used as greyscale image (Fig.12). A grayscale image as brightness component may also be obtained by a weighted colors conversion. Tables IV and V provide a performance comparison in term of directional correlation and randomness tests provided by *NPCR* and *UCAI* for "Lena" standard test image.

TABLE II: PLAIN IMAGE DIRECTIONAL AUTO-CORRELATION NORMALIZED COEFFICIENT

| Images | Horizontal | | | Vertical | | | Diagonal | | |
| --- | --- | --- | --- | --- | --- | --- | --- | --- | --- |
| | Red | Green | Blue | Red | Green | Blue | Red | Green | Blue |
| Lena-1024 | 0.9862 | 0.9824 | 0.9792 | 0.9723 | 0.9650 | 0.9599 | 0.9582 | 0.9482 | 0.9412 |
| Mushroom | 0.9794 | 0.9767 | 0.9613 | 0.9751 | 0.9710 | 0.9537 | 0.9597 | 0.9544 | 0.9298 |
| X-Ray | **0.9973** | **0.9967** | **0.9971** | **0.9976** | **0.9961** | **0.9967** | **0.9956** | **0.9947** | **0.9952** |
| Nature | 0.8594 | 0.8685 | 0.8431 | 0.8538 | 0.8647 | 0.8362 | 0.7983 | 0.8107 | 0.7751 |

TABLE III: CIPHER-IMAGE DIRECTIONAL AUTO-CORRELATION NORMALIZED COEFFICIENT

| Images | Horizontal | | | Vertical | | | Diagonal | | |
| --- | --- | --- | --- | --- | --- | --- | --- | --- | --- |
| | Red | Green | Blue | Red | Green | Blue | Red | Green | Blue |
| Lena-1024 | 0.00025 | -0.00059 | 0.004 | -0.0013 | -0.0016 | -0.0012 | 0.00002 | -0.00140 | 0.00060 |
| Mushroom | 0.00036 | 0.0074 | -0.0011 | 0.00510 | 0.0051 | -0.0016 | 0.00740 | -0.00430 | 0.00140 |
| X-Ray | 0.00100 | 0.0011 | 0.0008 | -0.0008 | -0.0008 | -0.0007 | 0.00000 | 0.00000 | 0.00003 |
| Nature | 0.00068 | -0.0017 | 0.0051 | -0.00420 | -0.0016 | 0.0040 | -0.00260 | -0.00032 | 0.00700 |

TABLE IV: CIPHERS COMAPAISON OF CORRELATION COEFFICIENTS FOR LENA-GREYSCALE IN FIG.2-A

| Algorithm | Horizontal | Vertical | Diagonal | Average |
| --- | --- | --- | --- | --- |
| None | 0.9792 | 0.9599 | 0.9412 | 0.9601 |
| **GFHT** | **0.0040** | **-0.0012** | **0.0006** | **0.0019** |
| Wang [12] | 0.0010 | 0.0022 | 0.0150 | 0.0061 |
| Wang [13] | 0.0021 | 0.0018 | 0.0014 | 0.0018 |
| Liu [16] | 0.0004 | 0.0021 | 0.0038 | 0.0021 |
| Wei [19] | 0.00062 | 0.0052 | 0.0069 | 0.0042 |

As confirmed by the results of the performance tests provided by table IV and V, the proposed GFHT cipher achieved high degree of security with characteristics similar to DNA coding and chaos theory logistic maps cryptography. To verify this assertion, the *UACI*, *NPCR* and average PCC provide values approaching theoretical limits of uniform white noise behavior for enciphered images applied for both color and greyscale images.

TABLE V: CIPHERS COMAPAISON FOR *NPCR* AND *UACI* FOR LENA-GREYSCALE

| Algorithm | NPCR (%) | UACI (%) |
| --- | --- | --- |
| **GFHT** | **99.50** | **33.45** |
| Wang [12] | 99.58 | 33.56 |
| Wang [13] | 99.65 | 33.48 |
| Liu [16] | 99.60 | 28.13 |
| Wei [19] | 99.21 | 33.28 |

The Blue layer taken as a grayscale was chosen since it provides similar auto-correlation coefficients comparted to those used for ciphers comparison in [12]. While GFHT cipher provides similar average characteristics as Wang et al. [13], it supersedes results provides by other works [12], [16], [19].

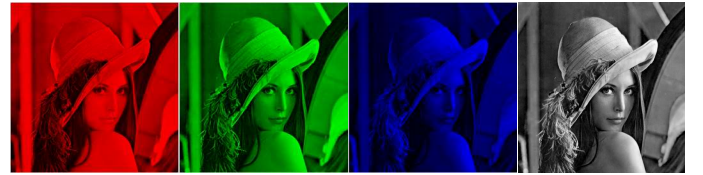

(a) Red layer  (b) Green layer  (c) Blue layer  (d) Grayscale

Fig. 12. RGB decomposition of "Lena-1024". The Blue layer is used as grayscale image for ciphers comparison in tables IV and V. Image resolution is reduced to 256×256 for comparison purpose.

### 4.5 Chi-Square, $\chi^2$, Hypotheses testing

Chi-square, $\chi^2$, is used to test hypotheses ($H_0|H_1$) about the observations to fit uniform distribution. The null hypothesis, $H_0$, is the assumption that the observed samples ($O_i$) are similar to the expected values ($E_i$), which manifests with $\chi^2 = 0$ with high cumulative density function (CDF), and the null hypothesis is accepted. In contrast, if the observed samples are different from the expected ones, the value of $\chi^2$ increases and $H_O$ is rejected if the *p*-value falls below the significance level which is the probability of rejecting the null hypothesis when it is true ($H_0$: $O_i=E_i$, $H_1$: $O_i≠E_i$). $\chi^2$ is based on the sums of squares of distributed variables ($O_i-E_i$) where the discrete statistic, is based on a finite number of samples, *K*, grouped into a set of *bins*. $\chi^2$ is given by the cumulative test statistic

$$\chi^2 = \sum_{i=1}^{K} \frac{(O_i - E_i)^2}{E_i} \qquad (27)$$

The larger the value of $\chi^2$, the more likely is that the distributions are significantly different. $p = \Pr[X \geq \chi^2]$ and the CDF with a degree of freedom (DOF), *v=bins-1*, is given by (28)

$$p = F(x,v) = \int_0^x \frac{t^{(v-2)/2}e^{-t/2}}{2^{v/2}\Gamma(v/2)} dt \tag{28}$$

where $\Gamma(.)$ is the Gamma function and Pr(.) is the probability.

To implement this statistical test for encrypted RGB layers, a sliding window of size $K$, of 300, 600, or 900 is used with three possible data scanline directions; horizontal, vertical or diagonal scanline (a.k.a. zigzag). The rate of *Goodness of Fit* ($R_{GOF}$) is the ratio of the number of accepted $H_0$ over the total number of tested sliding windows. Figure 13 illustrates a result of the $\chi^2$ hypotheses testing for "Lena-400×400", where $H_0$ rates, $R_{GOF}$, are 98.56%, 99.25%, 99.12% in horizontal, vertical and diagonal scanline respectively, for $K=600$, which validates the cipher as a scheme producing encrypted data adopting a uniform distribution. The figure also shows the normalized expected values (pixel's value/255) of encrypted pixels over the sliding windows indices. The number of sliding windows is: $N_w = 3.I_H I_W/r.K$, where $I_H$, $I_W$ and $r$ are image height, image width and window overlap ratio respectively. The calculated $H_0$ rejection rates (1-$R_{GOF}$) of 0.0075, 0.0088 and 0.0144 (from 1599 sliding window 23, 12 and 14 $H_0$ were rejected in horizontal, vertical and diagonal scanline respectively) agree with the rejection rates of pure uniformly distributed pseudo-random noise sequences.

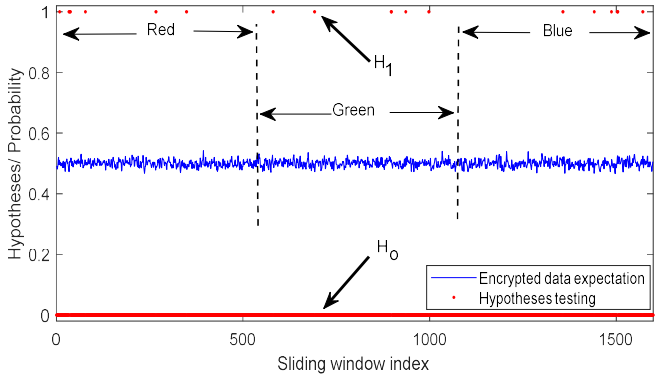

Fig. 13. Hypotheses testing using Chi-square, $\chi^2$, illustrating the uniform distribution of encrypted data. The RGB layers are concatenated.

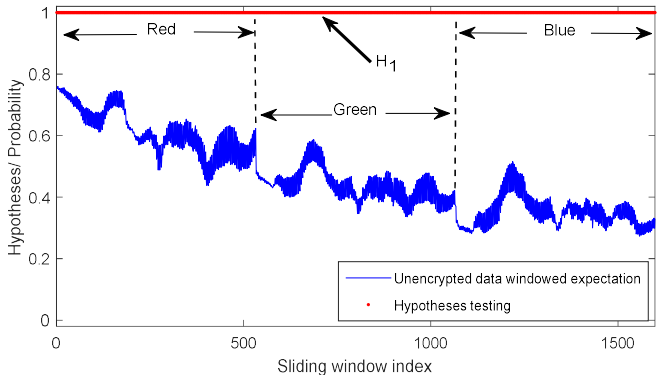

Fig. 14. Unencrypted data hypotheses testing failure using Chi-square, $\chi^2$.

The significance level is set to 0.01, the number of *bins* is 10 and DOF $v$ is 9; also, $v$ may be set as $v = \sqrt{K} + 1$ to be dynamically linked to the window size. Figure 14 shows $H_0$ rejection when processing the plain-image unencrypted data of "Lena-400" where $H_0$ is rejected with 100% for all sliding window indices, the processing is achieved over a total number of 1599 sliding window shifts with an overlap of 50% resulting in 533 windows for each of the Red, Green and Blue layers. All encrypted images of Fig.2, with resolutions: 1024×1024, 512×512, 400×400, 256×256 were tested for null hypothesis. Figure 15 shows the variation of the rate of the goodness of fit, $R_{GOF}$, for the 16 analyzed images (48 tests) in horizontal, vertical and diagonal scanlines; it shows the values of $R_{GOF}$ for each resolution. It is observed that the ripples depend on the ratio between image resolution and sliding window size. Increasing the image size decreases the ripples around the mean of $R_{GOF}$ given by 0.9872 for the set of 48 test. The $R_{GOF}$ means for the set of images grouped into resolutions of 256, 400, 512 and 1024 are 0.9879, 0.9878 0.9860 and 0.9872 respectively. The decrease of ripples with respect to image size is due to encryption being performed in a single block as given by the *P_Key* of (4); the characteristics are similar to those of pure uniform distribution white noise.

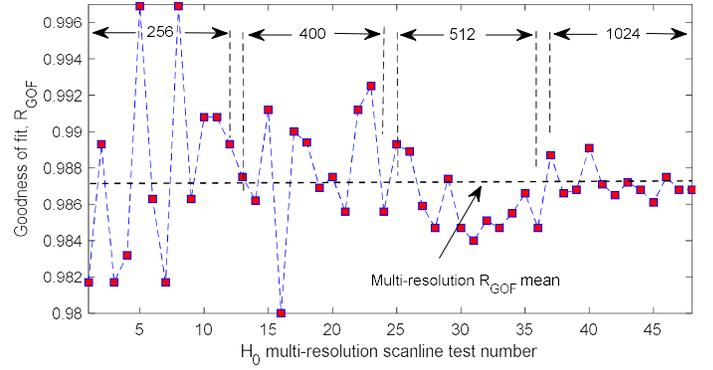

Fig. 15. Multi-resolution rate of goodness of fit, $R_{GOF}$, for tested images in horizontal, vertical and diagonal scanlines, $N_W$=600.

## V. CONCLUSION

In this paper, randomness evaluation of a symmetric genetic algorithm, GFHT, is presented. Pseudo-codes for both keys generation genetic algorithm implementation is presented first to show how such cipher is computationally efficient, afterwards, the signal processing approach for randomness evaluation is investigated based on random matrix theory circular law, power spectral density analysis, avalanche effect, 2-D autocorrelation, pixels randomness tests and chi-square, $\chi^2$, hypotheses testing. Results show that encrypted images adopt the statistical behavior of uniform white noise; hence validating the theoretical model by experimental testing. The GFHT algorithm, implementing an OTP key scheme, is used for image communication, suitable for secure communication over open networks. The performance comparison with chaos-genetic ciphers is also carried out.

## APPENDIX: UACI PROOF

Let *X* and *Y* be two uniformly distributed and independent variables, with their values in the interval [0, L]. The uniform probability density functions of *X*, *Y* are given by:

$$f_X(x) = \begin{cases} 1/L & \text{if } x \in [0, L] \\ 0 & \text{otherwise} \end{cases} \quad (29)$$

$$f_Y(y) = \begin{cases} 1/L & \text{if } y \in [0, L] \\ 0 & \text{otherwise} \end{cases} \quad (30)$$

The distance between the two variables $X$ and $Y$ distributed according to $f_X$ and $f_Y$ is a new random variable given by

$$Z = |Y - X| \quad (31)$$

therefore, the expected value of the new variable $Z$, is $\mathbb{E}(Z) = \mathbb{E}(|Y - X|)$, where the absolute value as a distance measurement is defined by

$$g(x, y) = |y - x| = \begin{cases} y - x & \text{if } y \geq x \\ x - y & \text{if } x \geq y \end{cases} \quad (32)$$

The joint probability density function, is the product of the probability density functions of $X$ and $Y$; given by $f_{XY}(x, y) = f_X(x)f_Y(y) = 1/L^2$ inside the squared space $[0, L] \times [0, L]$, where the expected value represents an area. Therefore, the expected value $\mathbb{E}(Z) = \mathbb{E}(g(x, y))$ is given by

$$\mathbb{E}(Z) = \int_0^L \int_0^L g(x,y) f_X(x) f_Y(y) dx dy$$
$$\mathbb{E}(Z) = \frac{1}{L^2} \int_0^L \int_0^L |x - y| dx dy$$
$$\mathbb{E}(Z) = \frac{1}{L^2} \left\{ \int_0^L \int_0^x (x-y) dy dx + \int_0^L \int_x^L (y-x) dy dx \right\}$$
$$\mathbb{E}(Z) = \frac{1}{L^2} \left( \frac{L^3}{6} + \frac{L^3}{6} \right) = \frac{L}{3} \quad (33)$$

with $L$=255, the distance, $Z$, expected value is $\mathbb{E}(Z) = 85$, while the normalized value represented by $L$=1, is $\mathbb{E}(Z) = 1/3$ or $\mathbb{E}(Z) = 33.33\%$.

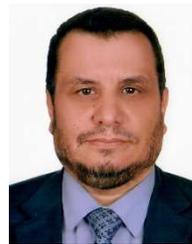

**Zoubir Hamici** (S'95–M'96–SM'09) was born in Barika, Algeria. He received the B.Eng. degree in electronics and control from the University of Annaba, Annaba, Algeria, in 1991, the D.E.A. (master's) degree in automation and signal processing from the Institut National Poly-technique de Lorraine (INPL), Nancy, France, in 1992, and the Ph.D. degree in signal processing from the University of Claude Bernard, Lyon I, France, in 1996. He is the Principal Investigator with the Communications and Embedded Signal Processing Research Laboratory, Faculty of Engineering and Technology at Al-Zaytoonah University of Jordan. Prof. Hamici current research interests include time-frequency mixed domain signal processing, intelligent learning control, optimization algorithms for electrical machines and wireless communications, embedded systems signal processing for wireless sensor networks and Industrial Internet of Things (IIoT), power systems cyber-physical systems and cybersecurity algorithms for cyber-physical systems.